\documentclass[11pt,letterpaper]{report}
\font\tenbf=cmbx10
\font\tenrm=cmr10
\font\tenit=cmti10
\font\elevenbf=cmbx10 scaled\magstep 1
\font\elevenrm=cmr10 scaled\magstep 1
\font\elevenit=cmti10 scaled\magstep 1

\font\ninerm=cmr9

\def\underl#1{$\underline{\smash{\hbox{#1}}}$}

\renewenvironment{thebibliography}[1]
 { \elevenrm 
   \begin{list}{\arabic{enumi}.}
    {\usecounter{enumi} \setlength{\parsep}{0pt}
     \setlength{\itemsep}{3pt} \settowidth{\labelwidth}{#1.}
     \sloppy
    }}{\end{list}}

\begin{document}
\begin{center}{{\elevenbf CAN REMOTE OBSERVING BE GOOD OBSERVING?\\
               \vglue 8pt
              REFLECTIONS ON PROCRUSTES AND ANTAEUS
\footnote {{\parindent 0pt 
\ninerm Slightly edited version of an article 
 published in 1993 in ``Observing at a Distance'', the 
proceedings of a workshop on Remote Observing, Eds: D.T.~Emerson \& 
R.G.~Clowes, World Scientific, p.~325.}} \\}}

\vglue 1.0cm
{\tenrm Felix J. Lockman 
\\}

\baselineskip=13pt
{\tenit National Radio Astronomy Observatory   \\}
\baselineskip=12pt
{\tenit Green Bank, WV 24944 USA}
\vglue 0.8cm
{\tenrm ABSTRACT}
\end{center}
\vglue 0.3cm
{\rightskip=3pc
 \leftskip=3pc
 \tenrm\baselineskip=12pt
 \noindent
Remote observing seeks to simulate the presence of the astronomer at the 
telescope.  While this is useful, and necessary in some circumstances, 
 simulation is not reality.  The
drive to abstract the astronomer from the instrument can have unpleasant
consequences, some of which are prefigured in the ancient tales of 
Procrustes and Antaeus.  This article, written in 1992 
for a conference proceedings on remote observing, 
is reprinted here with only slight editorial changes and the 
addition of a short Afterword.   I consider some of the human 
factors involved in remote observing, and suggest that our aim be to 
enhance rather than supplant the astronomer at the telescope.

\vglue 0.6cm}

{\elevenbf\noindent 1. Introduction: Why Are We Here?}
\vglue 0.3 cm

{\tenrm
\rightskip=3pc
 \leftskip=3pc
 \tenrm\baselineskip=12pt
 \noindent
``...we need to be reminded of the distinctions between the extraordinary 
power of science and the fallibility of those who practice it.''  
--  {\tenit A. Kornberg (1992)$^1$}
\vglue 0.3cm}

{\baselineskip=14pt
\elevenrm

  Most of this meeting has dealt with the interesting problems
that arise when we choose to operate a telescope remotely rather than
be there in person.  Of course, all space observatories must be run remotely,
and they have been reasonably successful, but only   at a cost that is 
 appalling  by terrestrial  standards.  If we decide
to support  remote observing at our ground-based telescopes because it is 
{\elevenit good}
rather than because it is  {\elevenit necessary}, we find ourselves 
asking questions like: What is good observing?  What is the output of
a successful observing run?  What is meant by ``high quality data''?
and so on.  To complicate matters, the increasing automation of 
observatory functions suggests to some that the 
local/remote distinction is
unimportant --  both observers sit in front of a computer and
communicate through it.  Can there any difference at all?

Let me approach the answers to these difficult and provocative questions by
posing yet another:  why  we are here?  That is, why are we here in person?
Why have so many of us faced the perils of airline food and maniacal baggage
handlers to sit in a room and hear each other talk?  Human speech has such a
low  data rate that it can be carried comfortably over mere
telephone lines, with ample  room  for illustrations.  
From the utilitarian view of information transfer our physical 
gathering together here is exceedingly strange.  Why didn't we convene 
remotely?  Is there some spooky  advantage to proximity over distance?

At NRAO we are building the Green Bank Telescope, which will be operated
as a visitor-oriented facility and will support remote as well as 
local observers.  We are supporting remote observing because some of our
users think that it will be a good feature, 
and because  it seems inexpensive.
But we're not doing it because it will produce better astronomy or better
astronomers.  No one knows.  And no one knows  if in the long
run it will be a blessing or a curse. 

To some, perhaps at older observatories, who routinely  struggle with balky,
uncontrollable  equipment in the depths of  longjohn winter nights, the notion
that automation could be overdone may seem ludicrous.  Even modern telescopes 
frequently suffer from computer systems that are a hopeless clutter of ad hoc 
languages and machines communicating with each other poorly, 
and with the hapless 
astronomer barely at all.  It might not seem credible that astronomy could be
done without an astronomer nearby.  But the trends to automation and 
abstraction
are firmly in place, and we need to decide how to use these tools before
the decision is made for us by default.

The ``reflections'' that follow are my  unsystematic comments
on these issues. The 
view is from the human, rather than  from the technical perspective, so 
 I will not discuss  what we {\elevenit can} do, but rather  
 what we {\elevenit should} do with our limited resources and time.
There is probably no observatory that couldn't perform better if it had more
money for instrumentation.  Should  these dollars be spent instead
on remote observing?
I believe that many of my comments are widely felt but rarely articulated.  
Some have already been expressed at this meeting, while others come from the 
 vast literature of applied
psychology, a  literature which includes many  ancient fables, for 
  the interaction of people with their creations has  always aroused
curiosity  and passion.

\vglue 0.6cm 
{\elevenbf\noindent 2.  Telescopes Produce Responsibility and Skills (and Data)
}
\vglue 0.4cm

We want our telescopes to produce good science, and good data are a part, 
but only a part,  of good science. 

\vglue 0.3cm

{\elevenit\noindent 2.1  Responsibility
}
\vglue 0.05cm
{\tenrm
\rightskip=3pc
 \leftskip=3pc
 \tenrm\baselineskip=12pt
 \noindent
 {\bf Responsible}. {\it adj.} \  {\bf 1}.~Legally 
or ethically accountable for the care or welfare of another.
{\bf 2}.~Involving personal accountability or ability to act
without guidance or superior authority. ... 
{\bf 4}.~Capable of making moral or rational decisions on one's 
own, and therefore answerable for one's behavior.
{\bf 5}.~Able to be trusted or depended upon.
{\it --  American Heritage Dictionary$^2$}
\vglue 0.3cm}

Observational astronomy is more than 
plucking  numbers from the sky --  we  also assign  a  reliability  to
those numbers.  When scientists  present the results of an experiment they
take responsibility for those results by attaching to them the 
most precious coin of the scientific realm: the individual scientist's
pledge to speak the truth.  

The definitions of the word ``responsible''  make it
clear that responsibility is something particular  to humans.  It makes
no sense to say that a computer is accountable for its own behavior, or that
it is capable of acting without superior authority.  
In society, responsibility is
so important  that we do not even credit it to  every person: 
a child or adolescent, no matter how intelligent, 
no matter how talented or athletically able,
might still not be  trusted to assume responsibility in some circumstances.

Who takes responsibility for data gathered remotely?  Who 
speaks for its correctness? Take the case of the 
Westerbork radio telescope.   Authors of 
a successful proposal to use this instrument are often not even notified 
  that the observations are being made.   Instead, 
sometime later they  receive in the mail  
a fully calibrated map of their field.   Who is responsible for the
correctness of these data?  Who knows how to interpret the lowest contour?  
Who should respond to a referee's comments?  
In this example, can the
remote astronomer be expected to vouch for the data quality?  But if
not the astronomer, then who?

This is an  extreme example of a split between the astronomer
who ends up with the data, and an observing specialist  who is responsible for 
what happens at the telescope.  Perhaps use of astronomical 
archives is also in this category.  The split is created naturally in 
most remote observing schemes.
I do not want to imply that an astronomer present at the 
telescope  automatically produces correct or responsible data,
or that the remote, but attentive astronomer must be utterly
irresponsible.  No, 
divided responsibility can exist even in locally operated systems, 
and to some extent is an inevitable outcome of the complexity of modern
observatories.   But much observing is done under less than perfect 
conditions, and proximity to the telescope 
increases the likelihood that variables of that particular day and time will be
scrutinized, that the weather will be noticed if 
it is unusual, and so on.    The more that one has the opportunity to 
 interact with every part of a telescope system,
 in some literal sense the closer one is to the 
telescope, the easier it is to be responsible for what occurs there.
\vglue 0.3cm

{\elevenit\noindent 2.2 Skills
}
\vglue 0.2cm

Anyone who spends  time around observatories 
knows that there are good observers and bad observers.  I 
believe that it is important to give our telescope users the opportunity
to become good observers.  Someone who actively plays with the 
equipment, tries out various combinations of things, and constantly
iterates on technique,  not only gets data and a sense of its 
correctness, but also develops skills which can make the next data  better.
The way to become a skilled observer is to participate in the 
observations as completely as possibly, to seek active control 
or understanding of every
phase of the process;  to try to recognize the difference between 
the basic limitations of an 
instrument and those limitations which are rooted in style or tradition.
It is hard to see how the consumer of Westerbork data described
above can even be called  an observer, much less a skilled one.  
And I find it difficult to believe that observing 
skills can be acquired more easily at a 
distance (if at all) than in close proximity to the telescope.

The drive for new instrumentation and new techniques usually
comes from the most skilled astronomers, for they are not only technically
competent, but motivated by specific but ever-changing astronomical interests.
They provide new ideas and the impetus to keep our telescopes 
at the state of the art.  We need skilled astronomers, not just
observing specialists, and 
we need to help more astronomers become skilled.  Students, especially,
 must learn about the instruments, 
and not simply about the control systems that are 
abstract representations of them.

\vskip 0.3cm

{\elevenit\noindent 2.3 Data
}
\vglue 0.2cm

Quite often the details really don't matter -- the 
routine data acquired from a telescope without special effort are good 
enough.  The  measurement of a high signal-to-noise spectrum, say, or 
of the flux density of an object, might require no subtlety at all, and
might as well be done in a batch process as by a skilled observer 
(provided that the sky is clear, etc.).  But I suspect that it
takes considerable experience to be able to 
decide that  an observation will be routine, and it may take 
several trips to the telescope, by someone, to set up and 
debug even a ``routine'' observing
procedure.

\vglue 0.6cm
{\elevenbf\noindent 3. Astronomers and Their Hammers
}
\vglue 0.3cm

{{\tenrm
\rightskip=3pc
 \leftskip=3pc
 \tenrm\baselineskip=12pt
 \noindent
``If you give a child a hammer, it will treat everything 
as if it were a nail.'' 

}\rightline{{\tenit -- attributed to Abraham Maslow (1908-1970)}}
\vglue 0.3 cm}

{\elevenit\noindent 3.1 Hammers Don't Gossip}
\vglue 0.1cm

  The Law of the Hammer, as stated by the  child 
psychologist Maslow, has general truth.
  We tend to define the possibilities for action 
by the tools we have at hand.  We tend to define our choices by the
options we are given.

A relevant example is a  menu-driven observing program.  It is 
a blessing for the novice who  immediately has a list of all 
possible choices and (via a click on a mouse) a way to 
effect a choice.  Menus are good organizational tools, they
enhance efficiency  and reduce 
errors.  But they are also dangerous for several reasons, and 
one is that they don't create, repeat or 
even understand gossip.   

The menu choices at an observatory pertain to real physical devices,
observing techniques, and so on.  
But most front-line observatories are 
 constantly improving all equipment 
 and increasing the understanding
and development of new techniques.  Also, things break.
While it may be possible in theory to keep 
a telescope 
control program always totally up-to-date, in practice it is rarely 
worthwhile to do so.  Instead, most observatories implicitly
rely  on local ``experts'' to
communicate the latest information.  Often,
the best expert is the person who worked the previous shift.

As an example, consider the VLA, which is so highly automated that it
is  run by batch processes  almost all
the time.  It is, however, in such a natural state of flux 
that inside information is frequently essential to get the best data.  
Visitors are assigned a local expert who advises on 
the most recent state of equipment, observing strategies, and data 
reduction schemes.  Even so, a VLA observing run is apt to start with 
the telescope operator's announcing that an antenna or two 
have been having some difficulties, 
 and have been offline for the last few hours.

Most of us have a  natural reluctance to commit opinions to writing 
or to a computer help file 
 unless they have been checked out and confirmed.  But often such 
raw hunches, rumors and gossip (``Receiver B seemed flaky to me.
I couldn't pin it down, and it looks ok now,  
but you may want to watch it.'') are the most
important bits of information an observer can get.  In automating telescopes,
we often try to remove the need for gossip because it is 
viewed as unnecessary, or even as evidence of a flaw in our 
system.  But we should face the fact that active telescope systems use 
 gossip and inside 
information to function efficiently, and that gossip evolves as
 the natural response to the need for extremely current though 
low-grade news.
Of course we want to purge obsolete information from our systems, but
why not also accept the role of gossip and plan to use it efficiently?
If there is somewhere a general-purpose telescope whose  
online documentation reflects the current status of the instrument, it is probably
a sign that that observatory has spent too much money on software and 
not enough on  new instrumentation.

\vglue 0.3cm
{\elevenit \noindent 3.2 But I Thought It Was a Nail!}
\vglue 0.3cm

A  more pernicious effect of the problem revealed by Maslow's 
 Law of the Hammer 
is that over time  observers will stop trying to do things 
that are not listed in the menu.   Menu-driven observing encourages
menu-driven thinking.
I am not talking about intimidation here
as much as laziness and ignorance, and the pressure is never
overt, but subtle and persistent, very persistent.
   We need to create a climate in which  astronomers 
imagine new uses of the instrument, combine existing functions in new 
ways, and tailor the equipment to the task, rather than the other way
around.  The imagination of an astronomer on-site is often stimulated 
just by seeing the telescope and its equipment, and recognizing that 
the current configuration  is merely one solution to a problem. 
We must encourage remote (as well as local) observers to break free of the
menus.

Maslow's Law reminds us of another interesting factor: astronomers may
use the available tools in ways that were not originally intended or
foreseen.  We will always be surprised, perhaps pleasantly, perhaps
not, by the consequences of human creativity and stubbornness.  

\vglue 0.3cm
{\elevenit \noindent 3.3 Well, You \underl{Should} Have Wanted a Hammer.}
\vglue 0.3cm
If remote observing comes to constitute a large fraction of the use of a 
telescope, I think that there will be a subtle, but tremendous, 
almost irresistible pressure to discourage equipment and observing methods that
don't easily lend themselves to control from a distance.  An example which is 
not as silly as it sounds is that there can never be an instrument on the 
Hubble Space Telescope that requires frequent tweaking by some technician  with a 
screwdriver.  Of course not: the HST has to run remotely.  But is it good to
be so restricted on our terrestrial telescopes?  That is, is it more 
important that  a new device be totally debugged, reliable, and integrated with the 
system before it gets on the telescope, or that it gets 
into use, producing results that may drive science in new directions, 
as quickly as possible?  I vote for speed.  If
we don't pay special attention, we may not even notice that we've created
 a restriction, a bed of Procrustes 
(see $\S 11$),  on which the unusual 
experiment won't fit, until all the creative users have gone elsewhere.

\vglue 0.3cm
{\elevenbf\noindent 4. The Mongoose and the Cobra (Science is Subtle)
}
\vglue 0.2cm

{\rightskip=3pc
 \leftskip=3pc
 \tenrm\baselineskip=12pt
 \noindent
``...though Rikki-tikki had never met a live cobra before, his
mother had fed him dead ones, and he knew that all a grown
mongoose's business in life was to fight and eat snakes.''
 {\tenit -- Rudyard Kipling (1893)$^3$}
\vglue 0.2cm}

{\rightskip=3pc
 \leftskip=3pc
 \tenrm\baselineskip=12pt
 \noindent
``In the vicissitudes of human experience and development, conflict
is an unfailing attribute.''
 {\tenit -- Jacob~Arlow (1985)$^4$}
\vglue 0.2cm
}

There is a fundamental tension in scientific activities that 
was alluded to 
in previous sections but here will be faced head-on.  Simply put, 
  the culture of science and the culture of management/administration
are not compatible.  Like the mongoose, who just can't 
abide the cobra, and the 
cobra, whose feelings are reciprocal, the tension is irreconcilable. 

A manager has to function smoothly and be reliable.  
That is the nature of the job. 
Schedules must be kept, payrolls met, the rent and electricity bill paid.
Major random events are definitely not welcome.  Scientists, on the other 
hand, like change.  Even though they need to be methodical and 
plan experiments well in advance, scientists still
 like change a lot.  They rarely do the same 
experiment over and over.  Asked ``what equipment will 
you need for your
research next year?'' they want to answer ``I can't tell.  It depends on what 
happens next month.''  Managers, understandably enough, find this attitude less
than helpful.  But if an organization is run for the ease of
 managers, science dies.  We've all heard about the U.S. Government 
agency that recently asked its
scientists to write a report giving their research plans for the next
year  {\elevenit including detailed information on all discoveries 
that would be made.}
But on the other hand, 
I'm not sure that anyone past adolescence would want to 
work for an organization that refused to act in a fairly 
predictable  fashion.

The point is that science is subtle.  
There are few reliable ways (other than 
benign neglect) to encourage it, but many ways to stifle it.  Our 
carefully developed 
 hardware and software tools, intended to enhance scientific
productivity,  can easily turn into  a
hammer thrust into 
 the hands of a scientist who very much needs a screwdriver. 
In particular, we at this
meeting are potentially very dangerous to science, for as scientists who
are involved in the way observatories are run, we can be both 
mongoose and cobra.
We have the ability to force our particular scientific style on our 
institutions, for better or worse, and may even do so unconsciously.

Software encapsulates the philosophies of its builders.  We need to 
remember that good science requires extreme flexibility of 
individual and institution, and we need to value and reward change
as much as we do order and efficiency.  
 We have to develop a culture of change in which  we 
become suspicious, and begin to worry,  
when things remain fixed for too long a time.

Which brings me to the question of how to measure the success of
 an observatory,
or the quality of science that it is doing.  This problem is so knotty that 
most everyone dodges it and concentrates on the 
more easily quantifiable: papers 
published, visitors serviced, or even more primitive issues like 
observing efficiency.  
I will not argue that telescope systems should be less than 
maximally efficient,
but efficiency itself is not a very selective 
criterion, for it says nothing about the 
quality of the data that are being gathered so efficiently, or the worth of 
the experiments themselves.  
A skilled observer puttering around a telescope, trying out
various schemes and just doing things differently, may discover
that the previous dozen observers who gathered their data with
high efficiency were getting the wrong numbers.
One characteristic of a good observatory may be 
that it lets its  users do 
{\elevenit exactly} what they want  for the largest 
possible fraction of the time, even if what they want to do seems 
horribly wrong or  utterly worthless!

\vglue 0.3cm
{\elevenbf\noindent 5.  The Importance of Screwing Up
}
\vglue 0.2cm
{\rightskip=3pc
 \leftskip=3pc
 \tenrm\baselineskip=12pt
 \noindent
{\tenit Another occult belief is that of being always right.  Never to 
err is only possible when insignificant questions are involved.  Those who 
attempt to achieve infallibility consistently will instead achieve essentially 
nothing at all.  In fact they should be reminded of the old wisdom that
``who does not know that he is a fool half of the time, is most certainly a 
fool all of the time''.  --Fritz Zwicky (1957)$^5$
}
\vglue 0.3cm}

It is important that the observer be able to make mistakes and
 to screw up an observation, whether
remotely or locally.  The feedback makes observers more skilled.
Also, it may reveal interesting features or possibilities of our 
systems that were not previously known.  But there is no need to belabor
the point.  Anyone who has ever worked as a scientist knows the basic truth of
Zwicky's comments.

We can permit astronomers to blunder around in our observatories, and to screw
up in various ways, only if we 
feel that they have some degree of responsibility.  I was very impressed on a 
visit to Yerkes Observatory to find that there never has been an interlock or
limit switch that prevents an observer from driving the 
movable floor into the back of 
the great 40-inch refractor.  As explained by my host, Lew Hobbs, the knowledge 
that such a catastrophe could occur through careless use 
of the controls they hold in their hands, has made the generations of
astronomers (and graduate students) who have worked at Yerkes 
more reliable than any mechanical device 
could be.  While I, personally, would
nonetheless install a limit switch, the general lesson is clear.
The good use of an instrument  requires some responsibility
on the part of the user.  The best way I know  to encourage that responsibility
is by the feedback that comes with 
proximity.  The feel of the Yerkes floor rising up, or the wry look on the 
telescope operator's face when asking an observer
 ``Do you really want to point at the ground?'' 
all heighten the observing experience, enhance observing skills, and contribute
to general attention to detail.  At the telescope, 
an error can produce more than an ``invalid operation'' 
return code on a TV screen.  The feedback is 
immediate and possibly profound.  Steel moves across the sky.
Alarms may sound.  Someone at this meeting said that a good remote
observing system should allow observers to break the equipment, and 
I agree, but only if they understand at all times that they have
that power, and will be able to  {\elevenit feel} the consequences.

\vglue 0.4cm
{\elevenbf\noindent 6. The Telescope Operator  -- Just Another Human}
\vglue 0.3cm

{\tenrm
\rightskip=3pc
 \leftskip=3pc
 \tenrm\baselineskip=12pt
 \noindent
``In spite of decades of effort and a huge investment of resources, modern
computers do not see very well.  In any ordinary sense of the word, they 
don't ``see'' at all. ... This blindness to ordinary environments contrasts
sharply with the success of machine recognition in artificial domains.  
Computers have long been able to recognize the magnetic letters stamped on 
checks; modern scanning systems can even ``read'' (i.e. identify the words
in) printed text.  It's only the real world that gives them trouble.''
-- {\tenit Ulric Neisser (1992)$^6$}
\vglue 0.3cm}

The principals in an observing session at the NRAO and many other facilities 
are the operator, whose 
main responsibility is to the telescope, and the observer, whose
responsibility is to the data. The primary characteristic of both observer and 
operator is that they are human beings.  
Humans do certain things very poorly, like stare at a control
panel waiting for something to happen, and other things very 
well, like respond creatively to unfamiliar  situations.
The issues of remote observing are not too far removed from those of 
remote operation, as both place distance between man and machine.  Much
has been written about remote operations in various industrial settings,
 for in some circumstances lives hang in the balance.

\vglue 0.3cm

{\tenrm
\rightskip=3pc
 \leftskip=3pc
 \tenrm\baselineskip=12pt
 \noindent
{\bf FIGHTING BOREDOM}

\noindent
FIGHTER PILOTS have G-LOC and transport pilots have B-LOC -- boredom-induced loss of
consciousness.  A Douglas Aircraft Co. test pilot says designers are contemplating 
installation of a beeper that would sound in the cockpit occasionally.  If the crew
does not respond by pressing a button or talking, an alarm would sound to rouse them.
--{\tenit Aviation Week and Space Technology (1991)$^7$}
\vglue 0.2cm}

At NRAO we have telescope operators because they improve 
overall observing efficiency.
One could always design a fully-automated 
system that responded to problems by going 
through an orderly shutdown and calling for help.  
The Hubble Space Telescope
does this all the time.  But more observing gets done if there is 
someone  on the spot to diagnose and perhaps repair  routine 
problems,  to help the observer if possible, to take over
certain operations so that the observer can get some sleep, 
 and to provide a high level 
of monitor and control of the entire telescope and its surroundings.  
Tom Ingerson, in his talk, 
noted that having a person as part of the telescope system can 
make a tremendous improvement in the operation of an 
observatory.  Human operators can be the best solution to the problem of how
to operate a telescope,
but only if we do  not expect them to act like machines.

The airline industry is quite familiar with this issue.  They joke 
 that the aircraft cockpit of the future will have only two
occupants -- a pilot and a large dog. The
dog is there to bite the
pilot if he tries to touch any  of the airplane's controls.  
And the pilot's job?
To feed the dog$^{8}$.
A similar message is given by the ``B-LOC'' article 
reproduced above.  The essence of the airline's problem is that 
computers can fly commercial airplanes better than humans 
{\elevenit provided that no unanticipated events occur}.  So the aviation 
companies face 
the dilemma of having to keep an aircraft crew 
ready to intervene, at a moment's 
notice, with their highest level of skill and concentration, 
in an environment which is mortally boring.  
Airbus, purveyors of perhaps the most automated commercial aircraft, 
holds ``overconfidence prevention classes'' to remind 
its pilots that
even if the computers say that everything is A-OK, 
the airplane may not be$^{8}$.

Many observatories are coming dangerously close to this state right now, 
and more will be so in the future.
The problem is not one that can be cured easily by
 motivational sessions with the telescope operator.  
Automation tends to 
reduce a person's   sense of responsibility because the computer
appears to have been given charge of the situation.  
At NRAO we recently debated whether there might be advantages to remote 
{\elevenit operations}  of the Green Bank Telescope,
i.e., to having the telescope operator located in the
 main lab building rather than at the telescope
several miles away.  After much thought, I am convinced that this
would be a grave error.  How can we ask someone to be responsible 
for a telescope and then place her miles away where she perceives it 
only at a substantial remove?

\vskip 0.5cm
{\tenrm
\rightskip=3pc
 \leftskip=3pc
 \tenrm\baselineskip=12pt
 \noindent
``I am what I am and that's all what I am .''
-- {\tenit Popeye the Sailor Man}
\vglue 0.2cm}

There are at least two problems with remote operations which also apply 
in some ways to remote observing.
  The first is reflected in the adage: 
out of sight, out of mind.  We really do forget about those things that do not
frequently intrude on our consciousness.    I was 
struck by Mel Wright's comment that he has found it necessary to cycle
staff from the Berkeley Campus to the Observatory at Hat Creek because people
get disconnected from what is going on if they are 
away from the telescope too long.   
A telescope operator who communicates with the telescope only 
through a computer
terminal is being taught an unconscious lesson every minute: that his 
main responsibility is 
to what is on the terminal; that the real telescope off yonder is
secondary, someone else's problem, or worse: an abstraction.  This is 
another manifestation of Maslow's Law.
The human tendency to put the ``out of sight out of mind'' 
should not be seen as a 
problem to be overcome by education or training; it is more like 
our inability to perceive X-rays than my inability to read Chinese -- 
the natural way of dealing with the immense flood of information presented
by the senses.  We block out what seems not immediately important.

The second problem is that we  make poor use of  peoples'
 talents if we  
restrict their information to what comes over computer terminals, video 
cameras, microphones, etc.,  unless it is absolutely necessary.
  There is  no adequate substitute for human
beings and their senses 
 when it comes to perception and creativity.
While it may seem that 
remotely controlled devices can convey information that is almost 
as good as one gets by  being there in person, 
serious efforts in this area have in practice been marked more by 
failure than by success$^{6}$.
   The video camera, e.g., is not a good substitute for the eye.\footnote{
{\ninerm  Again, I am aware that remote sensing is 
necessary in hostile environments like the core of a nuclear reactor
or the vacuum of space.  But I am concerned here with situations 
in which we have a choice, and the environment of most of our 
terrestrial telescopes is not especially forbidding.}}\ 
For one,  our eyes are constantly 
in motion, driven by a creativity and perception that 
cannot be replicated by 
machines.   If a telescope operator hears
a peculiar noise, he may want to touch a device, or smell it, or look around 
behind it, all without thinking especially hard or logically about the 
``search pattern'' he is going through.  It is not the same as peering
through a camera.
Our creative tendencies make it difficult for a machine to simulate a person.

Instead of replacing the human, perhaps we should replace the machine.
  Human beings are designed to be good at creative perception and 
decision making on the fly, so why waste 
that ability by restricting an operator 
to a desk behind a computer several miles away from the action,  
if he could be on the spot? I suggest that our goal should be to
use  remote sensing to supplement human perception, not to supplant it.

I do not know how to reconcile the desire for increased automation, and thus
increased reliability and control,
with the need to keep the operator fully involved with  the 
telescope.  The same considerations apply to the observer.
 Perhaps we should  not automate a task unless it produces an 
{\elevenbf overall} improvement in the telescope system, 
including  the functioning  of  operator and observer.  
This approach would deliberately leave certain jobs to be done by 
people, not machines.  
Perhaps we should give the operator other duties, 
 such as telescope maintenance, that fill the time and 
keep her fully in touch with the 
instrument.  Our observatories are run by 
and for human beings.  We should
 focus our efforts on using and enhancing human abilities, 
rather than attempting to simulate them.

\vglue 0.5cm
{\elevenbf\noindent 7.  TV is Only TV}
\vglue 0.2cm

{\tenrm
\rightskip=3pc
 \leftskip=3pc
 \tenrm\baselineskip=12pt
 \noindent
{\tenit March 28:}
``... we all landed safe \& sound, one of our servants (Margaret's maid)
excepted, who was very ill on the voyage \& is not expected to live ...''

\noindent
{\tenit June 2: }
``Rode over to the observatory in the dark to dinner and got beset by 
Van Renen's Dogs, a hungry savage pack of 5 or 6 large hounds \& curs
of low degree, who had all but eaten me \& my horse too ...''

\noindent
{\tenit July 30: }
``Rode over to Observatory Botanising \& shooting by the way. Shot a 
large Gull -- The Splendor of the flowers in the flats about the Camp
Ground is amazing.  Swept [the sky] till 3 AM.''
{\tenit { -- John Herschel (1834)$^9$}}
\vglue 0.3cm}

Astronomical expeditions of the past were heroic affairs, as much
a physical as a mental challenge.  At many observatories 
observing is still strenuous, though we are trying
 to change this.  But is it really wise to seek to 
turn experimental astronomy into an activity that is 
indistinguishable from watching TV?

Every once in a while it is  good  to remember that 
watching TV engages a very small part of the 
spectrum of human potential, and to first order,  the
content of the TV show is not important.  
In other words, it does not matter if we are watching a 
computer display or a 1970's sitcom;
while watching TV  we are essentially
at rest, using only our eyes and ears, 
focusing on a small part of our surroundings.
The flickering display induces a metabolic state that has 
been measured as less
energetic than sleep.  

To second order, there are  differences in muscular activity between 
using a computer and watching a TV show.  
The flick of the remote channel changer 
requires less coordination than light typing on a 
keyboard.  But the difference is still small on the 
scale of possible human activity.  

To third order, there is a qualitative difference
 in content of a TV screen and a 
computer screen, and  perhaps this is what 
distinguishes us from the majority of the U.S. population 
who spend a comparable amount of  time staring at a screen.  
Much creativity can occur in the interaction between programmer and 
computer, and I doubt that people go into computer science
because they like being paid to 
watch TV.  But that is what they end up doing, and
forcing others to do.  Perhaps it is not a good idea
to make astronomers and telescope operators dependent on such a 
narrow information channel.   I find it difficult to 
believe that human beings are designed to function well 
in so stupefying  an environment without major damage to the body and mind.

\vglue 0.6cm
{\elevenbf\noindent 8.  I Want to be Left Alone
}
\vglue 0.4cm

{\tenrm
\rightskip=3pc
 \leftskip=3pc
 \tenrm\baselineskip=12pt
 \noindent
``An observatory is stocked with human beings, after all, and
the isolation that dark, clear skies require is always a 
potential stress factor.  Some adapt, some crack, others 
just get grouchy.''  {\tenit -- Evens and Mulholland (1986)$^{10}$}
\vglue 0.3cm}

The astronomer at the telescope is blessedly free of most of life's 
distractions.  Usually, the phone does not  ring
constantly, few 
passers-by disrupt a contemplative moment, and even the choices for 
lunch are  limited.  Observatories give one the ability to 
concentrate on science.  This ability cannot be 
taken for granted for remote
observers, many of whom choose remote observing precisely because it allows
them to do other things simultaneously.  For this 
reason alone, it is  unlikely that the quality of remote observing
can ever be as high as that of local observing.  
In contrast, presence at the telescope encourages a high degree of 
involvement, especially if the astronomer has had to 
come some distance or does not use the instrument regularly.  

At this meeting there have been discussions 
 of remote observing centers:  communication and control facilities 
situated some distance from the telescope,
to which an astronomer would travel to observe.  
These could give one  isolation and 
focus, and also the context of gossip, and it seems possible
that with the proper
design and feedback such centers could  promote some responsibility for data 
and instrument.  What they do not give is familiarity with the 
instrument, or more knowledge of it or ability to manipulate it 
than was anticipated by those who wrote the communications programs.
Moreover, 
even at a remote observing center, can an astronomer who is unfamiliar
with the telescope ever really {\elevenit know} what is happening with it?

\pagebreak
\vglue 0.2cm
{\elevenbf\noindent 9.  Why We Should   Implement Remote Observing Anyway:
}
\vglue 0.4cm

1.  It gives us the ability to do short observing projects.  There are 
worthwhile, straightforward 
experiments that require use of a telescope for only a few hours, or,
in some cases, a few minutes.  Requiring the astronomer  to travel
to the telescope   produces a threshold for proposal submissions 
that is unlikely to be crossed for a brief observation.
 I believe that  many more short projects would be proposed if 
astronomers could be guaranteed a reasonable chance at getting 
good data without
having to travel  hours for just a few minutes of telescope time.  
Remote observing may even open up novel uses of telescopes, e.g., the 
monitoring of an object for just a few moments every evening.

2.  It increases the  efficiency of telescope use.  
If  some fraction of all proposals can be run remotely, telescope schedulers 
will be able to 
take advantage of particular weather conditions
or vagaries of the instrument.  
At the Green Bank Telescope, 
where some experiments will require a clear dry atmosphere while
others will do quite well in a steady rain, remote observing will 
allow us to match the program to the contingencies of the weather.

3.  It promotes new kinds of collaboration.  
At this meeting we have heard some amusing stories
of remote observers ``peering over the shoulder'' of a
collaborator who is at the telescope.  There are also other ways in 
which collaborations could evolve.  Imagine consulting a 
skilled specialist half way around the world for an opinion, 
or running several experiments simultaneously on geographically
separated instruments.  This latter feat is now done routinely in Very Long
Baseline Interferometry, but not without considerable help and
coordination at each telescope, and with little real-time feedback.
We can do better.

4.  It will open our instruments to more users.
The discipline required to produce a remotely operable
telescope system is liable
to  make that system more friendly and powerful for all users.
It may  open up a telescope for use beyond the  circle of insiders who
know the location of the  unlabeled switch that turns on
the receiver.  

5.  It may save  money?  There has been a strong sense at this meeting
that implementing a remote observing capability 
will not save an observatory any 
money.  This may be true if the telescope continues to be 
scheduled as it was before remote observing was possible.  If, 
however, as I suggest above, new observing forms evolve to take
advantage of the new capabilities, then the answer may be
different.  Large observing collaborations or short projects are 
now difficult to support because they are too expensive if everyone 
must come to the observatory.

6.  It is always worthwhile to add another tool to the toolbox.
It becomes a  problem only when, like Maslow's child with a hammer, 
we mistake the tool for the  task.

\pagebreak

\vglue 0.2cm
{\elevenbf\noindent 10. Why We Are Here
}
\vglue 0.4cm

By now, the analogy between our travel here to Tucson for this meeting
and an observer's travel to a telescope should be clear.  
A  meeting is much more than the
presentations.   Creative speculation, gossip, rumors and uncensored opinions
occupy the spaces between and around items in the formal agenda.  
Vital information is transferred  
in casual comments over coffee, comments that might never be made in an
open forum. In many cases we arrive at our personal evaluation of the 
formal presentations only after informal discussions.  People who are 
 separated from these interactions miss much of
the meeting, even if they absorb the complete agenda. 

At a meeting we also have great freedom of action, which
would not be the case if we were joined electronically.
If we wish, we can look not at the speaker, but at our neighbor who
may be frowning or laughing.   We are free to poke around the back of a 
computer, or to leave the room for a private conference
with one or two others.  This goes on all the time.  
In  {\elevenit The Wizard of Oz}, the  Wizard 
shouts, ``Pay no attention to the man behind the curtain'' as
his fraud is revealed by the dog who is not content with 
what is presented to him.  While we are at this meeting, we can
to some extent manage the flow of information on our own terms.
We can see for ourselves.
We are not bound by the menu.  

More subtle, but possibly more important in the long run, is that 
at this {\elevenit meeting} we are {\elevenit meeting} people.  
In normal contact we quickly develop a ``feel'' for the other
person, find out whether communication is easy or difficult, establish 
a level of trust or skepticism.  
At a good meeting, quite often a shared vocabulary is developed.
  We come to understand the meaning of
certain words by their context  and by watching the group react to their use.
Words, phrases, metaphors are created and spread throughout the 
community from the interactions at meetings.  
A good meeting conveys the same richness of experience as a good party, 
but usually with fewer morning-after regrets.

 There is also the matter of civility.  People linked electronically 
don't behave very well.  Anyone who has ever participated in 
 a ``phone'' meeting,
in which several groups communicate only by voice, knows
the poverty of these sessions and appreciates the value of proximity.  
In phone meetings I have  watched local folk 
read a newspaper, engage in a secondary conversation, or make faces or
gestures  when a remote party has the ``floor'' 
 --  all deadly insults if done in person and all expressing
a kind of hostility that I believe is  encouraged by the physical
absence, the abstraction,  
of our colleagues.  In my experience, phone meetings dehumanize interactions 
and, though often seemingly necessary and convenient, need to be supplemented
by frequent personal contact to avoid creating 
 major rifts in an organization.
A similar situation can arise in email.  Some colleagues whom I know to be 
warm, gentle and reasonably open-minded sound aggressive to the point of 
violence, wild, irresponsible and outright bigoted in their email.

Communication is never easy, even in person where we 
have access to the full range of the senses and plentiful redundancies of 
speech and gesture.  In the highly filtered world of telecommunications, it
is all the more difficult.
It seems that distance diminishes everything.
That's why we're here.

\vglue 0.6cm
{\elevenbf\noindent 11. Reflections on  Procrustes and Antaeus.
}
\vglue 0.4cm

{\tenrm
\rightskip=3pc
 \leftskip=3pc
 \tenrm\baselineskip=12pt
 \noindent
``Every time I come up with what seems like an original thought, 
it turns out that some damn Greek said it first.''
  {\tenit -- Anonymous}
\vglue 0.3cm}

{\elevenit\noindent  \underl{Procrustes}
}
\vglue 0.2cm

{{\tenrm
\rightskip=3pc
 \leftskip=3pc
 \tenrm\baselineskip=12pt
 \noindent
``I cannot stop to tell you hardly any of the adventures that
befell Theseus on the road to Athens.  It is enough to say
that he quite cleared that part of the country of the 
robbers, about whom King Pittheus had been so much alarmed.
One of these bad people was named Procrustes; and he was indeed 
a terrible fellow, and had an ugly way of making fun of 
poor travelers who happened to fall into his clutches.  In 
his cavern he had a bed, on which, with great pretense of 
hospitality, he invited his guests to lie down; but if they 
happened to be shorter than the bed, this wicked villain 
stretched them out by main force; or, if they were too tall, 
he lopped off their heads or feet, and laughed at what he
had done, as an excellent joke.  Thus, however weary a man 
might be, he never liked to lie in the bed of Procrustes.''
{\tenit -- Nathanial Hawthorne (1851)$^{11}$}
}}

\vglue 0.4cm
{\elevenit\noindent  \underl{Antaeus}
}
\vglue 0.2cm

{\tenrm
\rightskip=3pc
 \leftskip=3pc
 \tenrm\baselineskip=12pt
 \noindent
``There was one strange thing about Antaeus, of which I have not
yet told you .... 
 whenever this redoubtable Giant touched
the ground, either with his hand, his foot, or any other part of 
his body, he grew stronger than ever he had been before. 
The Earth, you remember, was his mother, and was very 
fond of him, as being almost the biggest of her children;
and so she took this method of keeping him always in full 
vigor.  Some persons affirm that he grew ten times stronger
at every touch; others say that it was only twice as strong.
But think of it!  Whenever Antaeus took a walk, supposing
it were but ten miles, and that he stepped a hundred yards
at a stride, you may try to cipher out how much mightier
he was, on sitting down again, than when he first started.''
{\tenit -- Nathanial Hawthorne (1851)$^{11}$}
}

\vglue 0.4cm
{\elevenit\noindent \underl{Astronomy}
}
\vglue 0.2cm

{\tenrm
\rightskip=3pc
 \leftskip=3pc
 \tenrm\baselineskip=12pt
 \noindent
``Such an attempt can take one of two antithetical forms: a search for
purity or a search for self-enlargement. ... The desire to purify oneself
is the desire to slim down, to peel away everything that is accidental, 
to will one thing, to intensify, to become a simpler and more transparent
being.  The desire to enlarge oneself is the desire to embrace more and
more possibilities, to be constantly learning, to give oneself over 
entirely to curiosity, to end by having envisaged all the possibilities 
of the past and of the future.''  -- {\tenit Richard Rorty (1986)$^{12}$}
}

\vglue 0.3cm

As individual scientists, we seek to reduce, isolate, extract,  the laws
from the tumultuous phenomena.  
Our accomplishments are judged on the narrowest of grounds.
But  our  scientific institutions, 
our observatories, even the way we approach our research, 
must be more expansive.

\pagebreak 

Several months ago the computer system that I had used for 
some years was shut down
and a modern workstation ``with 
100 times the power'' was placed on my desk.
In one blow all my programs were rendered inoperable, though admittedly the 
new computer could do wondrous things such as chirp like a cuckoo clock, 
display the latest weather map of the South Pacific, and produce a running 
graph  of the fraction of the CPU power  being used (which is 
usually near zero since it does not know how to run any of my programs).
  I accept its power, efficiency, and ultimate utility. 
I'm learning C++.   But when asked to ``name'' the 
machine so that it could be referenced  over the network, the choice 
was obvious.  It is called Procrustes.

Then there is Antaeus.  For us,
the most interesting part of the tale  occurs at his death 
 in battle with Hercules, who recognized that Antaeus, 
as the son of the earth, drew his strength from
frequent contact with it.  Every time 
Hercules flung Antaeus down, he rose renewed and stronger.

We find ourselves doing astronomy from  a variety of motives, but they are
usually positive:  we find it intellectually stimulating and 
a fascinating way to lead a life.  The challenges are not trivial.
Few of our colleagues have entered astronomy because they could not 
succeed  at the practice of law or made a mess at 
real estate sales, and so decided to settle for second best: the life of a 
scientist.  Observational astronomy has its own special attractions.
It can involve  elements of physics,  electronics, structural 
engineering, optics, mathematics,  computer science, photography,
 and may require luck and the ability to work  without sleep, 
alone in the cold.  Why would someone do such things?

Hercules defeated Antaeus by separating him from the source of his 
strength.  He held Antaeus up in the air away from 
 the ground, and Antaeus so removed was hardly Antaeus at all.  
His strength diminished, he became too weak to fight, was crushed, 
and  flung aside.  

We experimental scientists can also become intellectually 
weakened by distance, by separation -- 
transformed into consumers of 
data, not producers, and, like Antaeus, 
held away from the source of our strength, wither and fade. 
We are neither unprecedented nor unique in our situation, nor
in the struggle to prevent our tools  from diverting 
us from the reasons we made them in the first place.  

\vglue 0.5cm
{\elevenbf\noindent 12. Acknowledgments \hfil}
\vglue 0.4cm
{\baselineskip 14pt
I thank Cecil C.H.~Cullander, J.R.~(Rick) Fisher, and Dennis R.~Proffitt for 
many useful discussions on these issues, 
and Rick Fisher (again), Harvey S.~Liszt, Paul T.~Shannon, and especially 
Elizabeth B.~Lockman for comments that materially 
improved the manuscript.  The National Radio 
Astronomy Observatory is operated by Associated Universities, Inc., under a 
cooperative agreement with the National Science Foundation.
}

\pagebreak

{\elevenbf\noindent 13. References \hfil}
\vglue 0.2cm

\vglue 0.5cm 
{{\noindent \elevenbf 14. Afterword: June 2005}}
\vglue 0.2cm

\baselineskip 14pt
\elevenrm

	The paper reprinted here, at the persistent urging of Dr.~Paul 
Schechter, was published in a rather obscure proceedings of a small 
conference on remote observing held in 1992.   The meeting was abuzz 
with wonderful schemes for transmitting telescope control information 
over vast distances, but as the days progressed, it struck me that the 
priorities were being set not by the community of researchers -- 
the ostensible beneficiaries of all this work -- but by 
those interested in designing and implementing the systems.  So when 
I was asked on short notice to give one of the summary talks, I took 
the somewhat contrary, playful tone of suggesting that just because something 
{\elevenit could} be done did not mean that it 
{\elevenit should} be done without 
considering the consequences to creative research, or the cost compared to 
 other worthy uses of the same limited funds.  Not that I was 
opposed to modernizing telescope controls,  but neither was I convinced that 
remote observing was the straightest path to greater astronomical discovery.  

Talking was easy, but writing it up for the conference proceedings
 took real work, and a lot of reading.  
  With the help of acquaintances in other fields, 
and a bit of reckless disregard for my own professional reputation, 
this article emerged. The entire episode 
was challenging, nerve wracking, enlightening, and  
uncomfortably intimate.   Like most papers in conference proceedings, 
this one seemed destined to be forgotten, and in fact, 
to my knowledge it  has never been cited in another publication.  
Not even once.

And yet, {\elevenit Can Remote Observing be Good Observing: Reflections on 
Procrustes and Antaeus}  has had a  vigorous life and a wide circulation, 
an astonishingly wide circulation considering that for most years it has 
not been available in electronic form.  
I have received a regular stream of communications from people who stumble 
on it one way or another. Some whose opinion  I respect 
don't like it at all, but 
more typical is the report from a colleague  that copies of the paper  
appeared in every mailbox at her observatory at a time when it was 
considering a major reorganization.   Some  letters approach fan mail, 
which I am usually too embarrassed even to acknowledge,  and  every time I see 
Paul Schechter he makes me promise to make this paper 
available on astro-ph for wider distribution.  So here it is.

\vglue 0.4cm 
{{\elevenit\noindent \underl{Hands-Off Observing: 2005}}}
\vglue 0.2cm

Instead of posting this lightly edited reprint, I should have revised 
the paper entirely, for a lot has happened  since 1992 when the 
the paper was written and the Internet
was young.  The Green Bank Telescope is open for business, and 
``Doing Science Remotely'' is listed as one of 
{\elevenit Science} Magazine's ``Breakthroughs of the Year'' 
(Science 2004, 306, 2011).   There is the nascent 
National Virtual Observatory and a slew of robotic telescopes, 
while Westerbork -- my poster child for 
hands-off astronomy -- has been encouraging its users to become 
more involved with the instrument.   Blogs, chat rooms and wikkis offer 
new avenues for gossip.  Isolated graduate students, who spend their 
days `observing' from their Department basement, are now  trying to 
put together credible applications for postdoc positions.
And from the world of Sociology,  Reeves and Nass have  published 
their profound and wickedly funny studies on human-computer 
interactions, in a book ({\it CSLI Publications, 2002})
 which should be required reading for 
anyone who uses a computer or watches TV.  

 Parts of {\it Procrustes and Antaeus} now seem dated and even quaint, and 
its focus on  radio astronomy a bit parochial.  
Also, I have spent much of the last decade as a manager of an astronomical 
facility and have crossed and now recrossed the Mongoose-Cobra divide,  
gaining a new appreciation for the vital but conflicting roles of each 
species.    That adventure taught me that few things are cheaper than
 talk, and that you shouldn't bother to 
 take seriously the advice of anyone who won't make the effort 
to actually visit your facility.  As Texas Bix Bender says, 
``You can pretend to care.  You can't pretend to be there.'' 
In the last dozen years I have been asked several times to report on 
discoveries ``that will be made'' and  was  engaged in a struggle 
over remote telescope operations, which I lost.

So were I to begin this article afresh, it would certainly be
 different.  But lacking the time and interest right now to 
do the necessary scholarship, I have resisted the urge to revise 
except for minor tweaking of language. 
Anyway,  I suspect that a real revision won't be possible for 
another decade, when our 
profession is likely to be dominated by astronomers who have as little 
involvement in the acquisition of their data as they do in the harvesting 
of their food.   Shortly after writing this article, I listened to 
 a prominent astronomer describe his plan for a new, 
 state-of-the-art facility where  real astronomers would 
be kept safely outside the gates and the telescopes run by 
 observing specialists, who would give the 
astronomers the data they needed.   In 1993 that model sounded to me  
 like an  attempt by supercilious  managers to consolidate their power and 
stifle creativity, and it still does today, 
no matter how often it is repeated. 

`Hands-off'  observing may be part of a larger trend which goes by the 
buzzword `outsourcing':  the often-necessary delegation of 
responsibility to others for everything from auto repair and music 
making, to cooking and child rearing.  Why should astronomy 
buck this tide?  Or is it  `multitasking'?   Several colleagues have 
patiently explained to me that remote observing is the solution to the 
conflict they have between classroom responsibilities and 
research: one can graft an observing session onto the daily routine:  
teach by day and observe by night.   If they are correct, then 
what I initially suspected might be simple 
fear of travel turns out to be an even simpler 
alchemical desire to fabricate time!  

Of course, the issue is broader than 
remote vs.~local.  It could be described equally well as 
 hands-off vs.~hands-on; indifferent vs.~conscientious; 
distracted vs.~alert;  routine vs.~creative. 

\vskip 0.1in
{{\elevenit\noindent \underl{Observations}}}
\vskip 0.1in
	In anticipation of writing 
an eventual sequel to {\elevenit Procrustes and Antaeus}  
I often set aside relevant literature discovered in the course 
of other reading.   There's lots.  Here's a sample.

\baselineskip 12pt
\vskip 0.15in 
{\parindent 10pt 
{\tenrm
\rightskip=1pc
 \leftskip=1pc
In some ways modern science can be seen as the push to erase individual, 
craft skill from the scientific workplace, to ensure that no idiosyncratic 
local, tacit, or personal knowledge leaks into the product... Yet recent 
work in the sociology of science and engineering keeps discovering traces 
of craft in the modern scientific commodity.  Some lab technicians have 
``golden hands'' ...; some engineers are ``wizards'' ...; some physicists 
have ``physical intuition''.

\rightline{-- Susan Leigh Star in {\tenit The Right Tools for the Job, 
Princeton Univ.~Press, }}
\rightline{\tenit quoted in Science 1993, Vol.~260,  245.}}
\vskip 0.1in
\centerline{* * * * * *}
\vglue 0.1cm}

{\parindent 10pt
\tenrm
\rightskip=1pc
 \leftskip=1pc
``Normally, observing is a craft'', says Robert Kirshner of the 
Harvard-Smithsonian Astrophysical Observatory.  ``The way you learn is 
through trials,'' he says.  You figure out what you did wrong and try again.  
``With the Space Telescope you don't get any trials,''  he says.  If an 
observation goes wrong, an astronomer may have to wait months for a 
second chance.

...One suggestion that has come out..., he says, is that the 
[Space Telescope Science] 
Institute assign each outside observer a kind of caretaker and guide.  ``We
 need a person at the Institute responsible not only for the bureaucratic 
paper maze but also the meaning of the observations,'' says Kirshner -- 
someone who understands the ultimate goals of a project and makes sure the 
researchers get the data they need. 

\rightline{ -- Faye Flam in {\tenit 
Space Telescope Institute: Inside the Black Box, Science 1993, Vol.~260, 1716.}}
\vglue 0.1cm
\vskip 0.1in
\centerline{* * * * * *}
\vglue 0.1cm}

{\tenrm \parindent 10pt 
\rightskip=1pc
 \leftskip=1pc
{\tenbf Look at the raw data.}  There is no substitute for viewing the 
data at first hand.  Take a seat at the bedside and interview the
patient yourself; watch the oscilloscope trace; inspect the gel 
while still wet.  Of course, there is no question that further 
processing of data is essential for their management, analysis 
and presentation.  The problem is that most of us don't really 
understand how automated packaging tools work.  Looking at the 
raw data provides a check against the automatic averaging of 
unusual, subtle, or contradictory phenomena.

\rightline{-- D.~Paydarfar \& W.J.~Schwartz {\tenit An Algorithm for 
Discovery, Science 2001, Vol.~292, 13.}}
\vskip 0.1in
\centerline{* * * * * *}
\vglue 0.1cm}

\vskip 0.04in
{\tenrm \parindent 10pt 
\rightskip=1pc
 \leftskip=1pc
   I loved reading about these experiments and tried repeating some of 
them for myself --- our Hoover was a good substitute for Boyle's air pump.  
I loved the playfulness of the whole book, so different from the 
philosophical dialogs in 
{\tenit The Sceptical Chymist}.  (Indeed, Boyle himself 
was not unaware of this:  ``I distain not to take notice even of ludicrous 
experiments, and think that the plays of boys may sometimes deserve to 
be the study of philosophers'').

\rightline{--  O.~Sacks 2001, in {\tenit Uncle Tungsten: Memories of a 
Chemical Boyhood; Knopf.}}
\vskip 0.1in
\centerline{* * * * * *}
\vglue 0.1cm}

\vskip 0.1in
{\tenrm \parindent 10pt 
\rightskip=1pc
 \leftskip=1pc
...children discover new thinking strategies while succeeding at a 
task, as well as while failing at it... children who come up with 
several problem-solving strategies, even wrong ones, frequently learn 
more than those who generate just one or two strategies, even correct ones.

\rightline{-- B.~Bower in {\tenit Science News 2001, Vol.~159, 172.}}
\vskip 0.1in
\centerline{* * * * * *}
\vglue 0.1cm}

\vskip 0.04in
{\tenrm  \parindent 10pt 
\rightskip=1pc
 \leftskip=1pc
The content of life is limited by the amount of information we 
can process through attention.  In this sense attention, or psychic 
energy, is our most scarce resource.

\rightline{-- M.~Csikszentmihalyi 2004 in 
{\tenit Psychology and Consumer Culture,}}
\rightline{\tenit   American Psychological Assn.~p.~91.}
\vskip 0.1in
\centerline{* * * * * *}
\vglue 0.1cm}

\vskip 0.05in
{\tenrm \parindent 10pt 
\rightskip=1pc
 \leftskip=1pc
By tradition, impatience is a vice.  Haste makes waste.  Even if our 
technological world seems inspired by the modernist calculations of 
Benjamin Franklin, we can all think of a few remaining human activities 
that cannot profitably be rushed.  ``There are two cardinal sins,'' Kafka 
said, ``from which all the others spring: impatience and laziness.''  
There's the paradox --- maybe it's laziness, not industriousness, when 
we succumb to the economics of time.

\rightline{   -- James Gleick 1999, in {\tenit  
Faster: The Acceleration of Just about Everything, Pantheon.}}
\vskip 0.1in
\centerline{* * * * * *}
\vglue 0.1cm}

\vskip 0.1in
{\tenrm \parindent 10pt 
\rightskip=1pc
 \leftskip=1pc
 I actually removed a book from my syllabus last year because
I couldn't figure out how to PowerPoint it... 
When I read this book I thought, my head's filled with ideas, and 
now I've got to write out exactly what those ideas are, and --- they're 
not neat.

\rightline{ -- C.~Nass quoted by Ian Parker in {\tenit The New Yorker, 
May 28, 2004, p 76.}}
\vskip 0.1in
\centerline{* * * * * *}
\vglue 0.1cm}

\vskip 0.05in
{\tenrm \parindent 10pt 
\rightskip=1pc
 \leftskip=1pc
	Bullet outlines dilute thought. 
{\ \ \ \ \ \ \ -- 
E.R. Tufte 2004, in {\tenit The Cognitive Style of Powerpoint}}
\vskip 0.1in
\centerline{* * * * * *}
\vglue 0.1cm}

\vskip 0.1in
{\tenrm \parindent 10pt 
\rightskip=1pc
 \leftskip=1pc
 [a collaboratory is] a virtual entity created by means of a 
computer network...  A collaboratory cannot 
replace the richness or the commitment engendered by face-to-face 
interactions.  As in other collaborative arrangements, concerns 
surrounding trust, motivation, and normative practice for data access, 
ownership, and acknowledgment can hinder collaboratory function. 
\rightline{
S.~Teasley and S.~Wolinsky, {\tenit  Scientific Collaborations at a Distance,}}
\rightline{\tenit   Science 2001, Vol.~292, 2254.}
\vskip 0.1in
\centerline{* * * * * *}
\vglue 0.1cm}

\vskip 0.1in
{\tenrm \parindent 10pt 
\rightskip=1pc
 \leftskip=1pc
What seems most obvious is that 
 media are {\tenit tools}, pieces of hardware, not players in social 
life.  Like all other tools, it seems that media simply help people 
accommodate tasks, learn new information, or entertain themselves.  
People don't have social relationships with tools... 
We now think our intuitions were wrong, however.
 People respond socially and naturally 
to media even though they believe it is not reasonable to do so, 
and even though they don't think that these responses characterize 
themselves...  

...people are not evolved to twentieth-century technology.  The 
human brain evolved in a world in which {\it only} humans exhibited 
rich social behavior, and a world in which {\it all} perceived 
objects were real physical objects.  Anything that {\it seemed} 
to be a real person or place {\it was} real.

People have done amazing things in our labs.  They have taken great 
care not to make a computer feel bad, they've felt physically 
threatened by mere pictures, and they've attributed to an animated 
line drawing a personality as rich as that of their best friend.  It
eventually occurred to us that people were not doing these things 
because they were childish, inexperienced, distracted or because they 
needed a metaphor.  We had to acknowledge that these responses were 
fundamentally human, and we had to acknowledge that they were important.

Psychologically, the PC is not terribly different from the TV... 
Claims about amplified 
responses to new media are often exaggerated.  Our research is a reminder 
that we can cry when we read, and we can be bored in a virtual 
world.  Social and natural responses come from people, not from media 
themselves.  Ultimately, it's the pictures in our heads that matter, 
not the ones on the screen.  

\rightline{ -- Byron Reeves \& Clifford Nass 2002 in }
\rightline{{ \tenit The Media Equation}, CSLI Publications}
\vskip 0.1in
\centerline{* * * * * *}
\vglue 0.1cm}

\vskip 0.1in
{\tenrm \parindent 10pt 
\rightskip=1pc
 \leftskip=1pc
  Boy, for a Nobel Prize winner, Philip Anderson doesn't know much about 
experiments... he pooh-poohs the notion that thought processes can 
interfere with physics.  Any experimentalist could have told him that 
he is wrong.  If you turn your back on an electronic counter, it will 
certainly start counting backwards.  If you go to the bathroom, the 
temperature regulator is sure to fail.  If you start a scan and then go 
to lunch, the stepper motor will invariably jam just after the door closes.  
And if you should dare to take a vacation, there is no limit to the 
disasters that can happen. 
{\hfill -- P.~Kolodner  in {\tenit Physics Today, October 1991, p.~146}}
\vskip 0.1in
\centerline{* * * * * *}
\vglue 0.1cm}

\vskip 0.1in
{\tenrm \parindent 10pt 
\rightskip=1pc
 \leftskip=1pc
A couple of years ago I moved my big 
{\tenit Random House Unabridged Dictionary} 
from a shelf to a table that I pass a dozen times a day.  Within days, 
I was using the dictionary more frequently than before.  
On a long vacation ... I observed that the act of moving my bike 
from the garage where I ``safely'' kept it (rusting) to the back steps -- 
a move of about 10 yards -- caused usage to shoot up instantly.  Why? 
Being there.

\rightline{-- Tom Peters, in {\tenit Forbes ASP June 3 1996, 148.}}
\vskip 0.2in
\centerline{* * * * * *}
\vglue 0.1cm}

\vskip 0.1in
{\tenrm \parindent 10pt 
\rightskip=1pc
 \leftskip=1pc
You can pretend to care.  You can't pretend to be there.

\rightline{--  Texas Bix Bender {\tenit (quoted in Tom Peters 1996)}}
}

\end{document}